\documentclass{moriond}
\usepackage{amsmath,bm}
\usepackage{amssymb}

\def\st{\scriptstyle}

\def\be{\begin{equation}}
\def\ee{\end{equation}}
\def\bea{\begin{eqnarray}}
\def\eea{\end{eqnarray}}

\def\eq{\mathrm{eq}}
\def\stop{\widetilde t}

\newcommand{\Photo}{}

\begin{document}
\vspace*{4cm}
\title{DARK MATTER AND LONG-LIVED PARTICLES AT THE LHC}

\author{JAN HEISIG}

\address{Institute for Theoretical Particle Physics and Cosmology, RWTH Aachen University,\\ Sommerfeldstra\ss e 16, D-52056 Aachen, Germany}

\maketitle\abstracts{
While the paradigm of a weakly interacting massive particle (WIMP) has guided our search strategies for dark matter in the past decades, their null-results have stimulated growing interest in alternative explanations pointing towards non-standard signatures. 
In this article we discuss the phenomenology of dark matter models that predict long-lived particle at the LHC\@.
We focus on models with a $Z_2$-odd dark sector where -- in decreasing order of the dark matter coupling -- a coannihilation, conversion-driven freeze-out or superWIMP/freeze-in scenario could be realized. 
}

\section{Introduction}

Pinpointing the nature of dark matter is among the key scientific goals of the LHC\@. So far searches for dark matter have strongly focussed on missing energy signatures following the widely studied paradigm of WIMP dark matter. However, the LHC has not found any hint for a corresponding signal yet, proceeding to strengthen the constraints on WIMP dark matter models. At the same time, if the WIMP scenario is not realized in nature, a potential signal of dark matter related physics might hide in other places. It is therefore of utmost importance to investigate alternative ideas of dark matter genesis that could point towards new signatures. 

Motivated by a variety of theories beyond the standard model long-lived particle (LLP) searches have recently attracted growing interest. LLPs provide a wide range of possible signatures: highly ionizing, disappearing or kinked tracks (charged LLPs), displaced vertices (charged or neutral LLPs) as well as trackless jets or displaced leptons (neutral LLPs). While many LLP signatures provide very promising search prospects exploiting extremely low (and often solely instrumental) backgrounds they are typically hard to trigger on. Specialized trigger settings -- often relying on additional activity in the event -- are needed. This makes it a timely enterprise to investigate a comprehensive search program for LLPs in order to fully exploit the immense capability of the LHC to illuminate physics beyond the standard model.

This article constitutes a contribution to this effort by investigating dark matter scenarios that predict LLP signatures at the LHC\@. 
We focus on models with a $Z_2$-odd dark sector. 
The prime example is the well-known coannihilation scenario\,\cite{Griest:1990kh,Edsjo:1997bg} in which the coannihilating particle might or might not appear long-lived at collider time scales depending on the mass splitting between the coannihilating partner and dark matter.
While the canonical coannihilation scenario assumes relative chemical equilibrium in the dark sector during dark matter freeze-out
other viable possibilities exist with couplings significantly weaker than the weak force. In conversion-driven freeze-out\,\cite{Garny:2017rxs,Garny:2018icg} (or co-scattering\,\cite{DAgnolo:2017dbv}) the decoupling from relative chemical equilibrium (facilitated by conversion processes) governs the dark matter abundance. For even smaller dark matter couplings chemical equilibrium of dark matter might never even have been established leading to a superWIMP\,\cite{Feng:2003uy} (or freeze-in\,\cite{Hall:2009bx}) scenario.
Interestingly, for particles in the GeV to TeV range a departure from relative chemical equilibrium during freeze-out implies macroscopic decay length at the LHC -- an intriguing coincidence that renders the LHC to be a powerful tool to explore these scenarios. 
We discuss the appearance of LLPs in coannihilation and conversion-driven freeze-out scenarios in Sec.~\ref{sec:cdfo} while focussing on a concrete realization of the latter in Sec.~\ref{sec:rocdfo}. In Sec.~\ref{sec:sW} we comment on thermally decoupled dark matter before finally concluding in Sec.~\ref{sec:Con}.

\section{From coannihilation to conversion-driven freeze-out}\label{sec:cdfo}

A $Z_2$ symmetry (or a larger symmetry with a $Z_2$ subgroup) is commonly imposed in theories beyond the standard model in order to stabilize dark matter. UV-complete models often come with an entire $Z_2$-odd sector. This opens up the possibility for interesting phenomena regarding the evolution of its particle densities in the early Universe. 
A prime example in this concern is the well-know coannihilation scenario.\cite{Griest:1990kh,Edsjo:1997bg,Baker:2015qna} For small relative mass splittings between a 
heavier $Z_2$-odd state, $\chi_2$, and dark matter, $\chi_1$, $\Delta m/m_{\chi_1} \equiv (m_{\chi_2}-m_{\chi_1})/m_{\chi_1} \lesssim 10 \%$,\footnote{The exact value depends on the hierarchy between the involved couplings. In extreme cases coannihilation can be important for much larger mass splittings.\cite{DAgnolo:2018wcn}} the relative number density of the heavier state could still be significant during dark matter freeze-out:
\begin{equation}
n_{\chi_2}^\eq/n_{\chi_1}^\eq \propto \mathrm{e}^{-\Delta m /T_\mathrm{f}}\simeq \mathrm{e}^{-25\, \Delta m /m_{\chi_1}}\,.
\end{equation}
In the last expression we inserted $T_\mathrm{f}\simeq m_{\chi_1}/25$ as the typical freeze-out temperature. 
Consequently, $\chi_2$ participates in the freeze-out processes providing additional annihilation channels that can deplete the number density in the dark sector and, hence, due to conversion processes within the dark sector, the number density of dark matter.

\subsection{Coannihilation and long-lived particles}\label{sec:coannLLP}

Since the $Z_2$ symmetry forces the heavier states to decay into lighter states of the dark sector, 
a small relative mass splitting potentially leads to a kinematic suppression of the decay width. 
Therefore coannihilation scenarios can provide LLPs\@.  A prominent example is the stau coannihilation strip of the constrained MSSM\,\footnote{Minimal supersymmetric standard model.}, parts of which are most strongly constrained by searches for heavy stable charged particles.\cite{Citron:2012fg,Desai:2014uha,Heisig:2015yla} Another example concerns minimal dark matter\,\cite{Cirelli:2005uq} 
which extends the standard model by an electroweak multiplet. Here, a naturally small mass splitting between the neutral and charges states of the multiplet (of order $100\,\mathrm{MeV}$) arises from electroweak corrections. For the fermion triplet, which corresponds to a wino dark matter scenario in supersymmetry, the strongest LHC constraints arise from searches for disappearing tracks excluding masses up to 430\,GeV.\cite{ATLAS:2017bna} However, so far much stronger limits arise from indirect detection experiments.\cite{Cuoco:2017iax} 
Other coannihilation scenarios that naturally provide LLPs are \emph{e.g.}~pseudo-Dirac dark matter models\,\cite{DeSimone:2010tf,Davoli:2017swj} and models with colored dark sectors.\cite{Ellis:2015vaa,ElHedri:2017nny,Davoli:2018mau}

\subsection{Parasitical dark matter}\label{sec:LLPmir}

In the standard treatment of coannihilation conversion processes within the dark sector
are assumed to be efficient during freeze-out and relative chemical equilibrium is maintained, $n_{\chi_i}/n_{\chi_i}^{\eq}=n_{\chi_j}/n_{\chi_j}^{\eq}$.\footnote{This assumption is commonly made in
numerical relic density calculators.\cite{Belanger:2018ccd,Ambrogi:2018jqj,Bringmann:2018lay}}
 In this case annihilations in the dark sector can be described by an effective, thermally averaged cross section\,\cite{Edsjo:1997bg}
\begin{equation}
\langle\sigma v\rangle_\text{eff} = \sum_{i,j}\langle\sigma v\rangle_{ij}\frac{n_{\chi_i}^\eq}{n^\eq}\frac{n_{\chi_j}^\eq}{n^\eq}\,,
\end{equation}
where $n^\eq=\sum_i n_i^\eq$. For very small couplings of the dark matter to the standard model $\langle\sigma v\rangle_{ij}$
can be negligible for all channels containing dark matter in the initial state. The dilution of the number density of the dark sector is then driven entirely by annihilations of heavier states and not by dark matter annihilations. In this case the relic density becomes independent of the coupling strength of dark matter. However, this conclusion is only true for couplings that are still large enough to maintain relative chemical equilibrium.\footnote{Note that conversion rates are enhance compared to annihilations by a Boltzmann factor of order $\mathrm{e}^{m_\chi/T}$.} For even smaller couplings relative chemical equilibrium breaks down. In this case conversion processes are responsible for the chemical decoupling of dark matter and hence set the relic density. This \emph{conversion-driven freeze-out} mechanism is phenomenologically distinct and opens up a new region in parameter space where coannihilation would lead to under-abundant dark matter, if relative chemical equilibrium would hold.

\subsection{The ``LLP miracle''}\label{sec:LLPmir}

\begin{figure}[t]
\centering
\setlength{\unitlength}{1\textwidth}
\begin{picture}(0.47,0.25)
\put(0.0,-0.01){\includegraphics[width=0.38\linewidth]{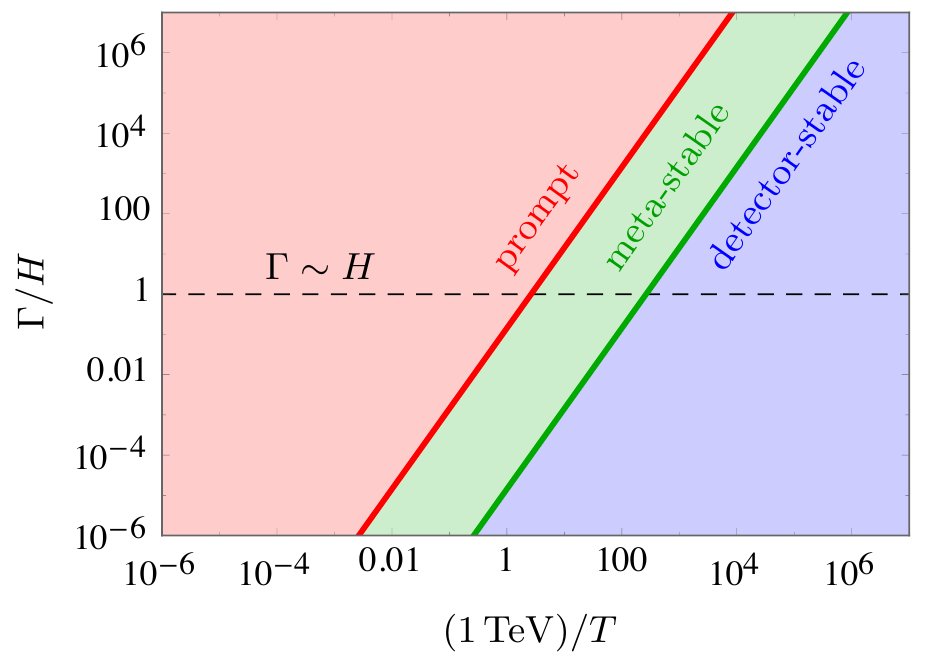}}
\end{picture}
\caption{Ratio between decay rate and Hubble rate as a function of the inverse temperature.}
\label{fig:rates}
\end{figure}

The departure from relative chemical equilibrium has an immediate consequence for the possible decay length of the heavier states. As the decay contributes to the conversions, requiring their rate to become inefficient necessarily requires 
\begin{equation}
\label{eq:decH}
\Gamma_\mathrm{dec}\lesssim H\,. 
\end{equation}
In the radiation dominated Universe $H = \sqrt{g_*/90}\, \pi T^2/M_\mathrm{Pl}$, where $M_\mathrm{Pl}\simeq 2.44\times 10^{18}\,\mathrm{GeV}$ is the reduced Planck mass. We can translates the inverse Hubble rate into a length. Using $g_*=100$, the inequality \eqref{eq:decH} then reads
\begin{equation}
c\tau\gtrsim H^{-1} \simeq1.5\,\mathrm{cm} \left(\frac{(100\,\mathrm{GeV})^2}{T^2}\right)\,.
\end{equation}
This is an important results which states that for particles in the GeV to TeV range a departure from relative chemical equilibrium during freeze-out ($T\simeq m_\chi/30$) implies macroscopic decay length at the LHC -- an intriguing coincidence that renders the LHC to be a powerful tool to explore these scenarios. Figure~\ref{fig:rates} illustrates the prompt, meta-stable and detector-stable regime in the plane spanned by the inverse temperature and $\Gamma_\mathrm{dec}/H$.

\section{Realizations of conversion-driven freeze-out}\label{sec:rocdfo}

In this section we discuss a realization of conversion-driven freeze-out within a simplified dark matter model.
We consider an extension of 
the standard model by a neutral Majorana fermion $\chi$ 
and a colored scalar particle $\tilde q$ that acts as a ($t$-channel) mediator of the dark matter interactions
with the standard model quarks $q$:
\begin{equation}
    \mathcal{L}_\text{int} = |D_\mu \tilde q|^2 + \lambda_\chi \tilde q\, \bar{q}\,\frac{1-\gamma_5}{2}\chi +\text{h.c.}\,.
    \label{eq:tchmodel}
\end{equation}
Here $D_\mu$ is the usual covariant derivative and $\lambda_\chi$ is a coupling strength of the dark matter interaction. For a certain choice of $\lambda_\chi$ this model resembles a subset of the MSSM, while smaller couplings could be realized in extensions of the MSSM.\cite{Belanger:2005kh} However, we do not refer to any particular UV-complete theory here considering $\lambda_\chi$ as a free parameter. 

Imposing a $Z_2$ symmetry under which all standard model particles are even
while $\chi\to-\chi$ and $\tilde q\to -\tilde q$ are odd, the Majorana fermion $\chi$
provides a viable dark matter candidate for $m_\chi<m_{\tilde q}$. 
We consider the cases of a bottom- and top-philic model, $q=b, t$, providing a distinct phenomenology. While the mass of the bottom is mostly small compared to the energies of the relevant processes the mass of the top is sizable leading to additional suppressions \emph{e.g.}~of the mediator decay.

\subsection{Cosmologically viable solutions}

\begin{figure}[t]
    \centering
    \setlength{\unitlength}{1\linewidth}
        \begin{picture}(1,0.44)
\put(0.015,-0.02){\includegraphics[width=0.47\textwidth]{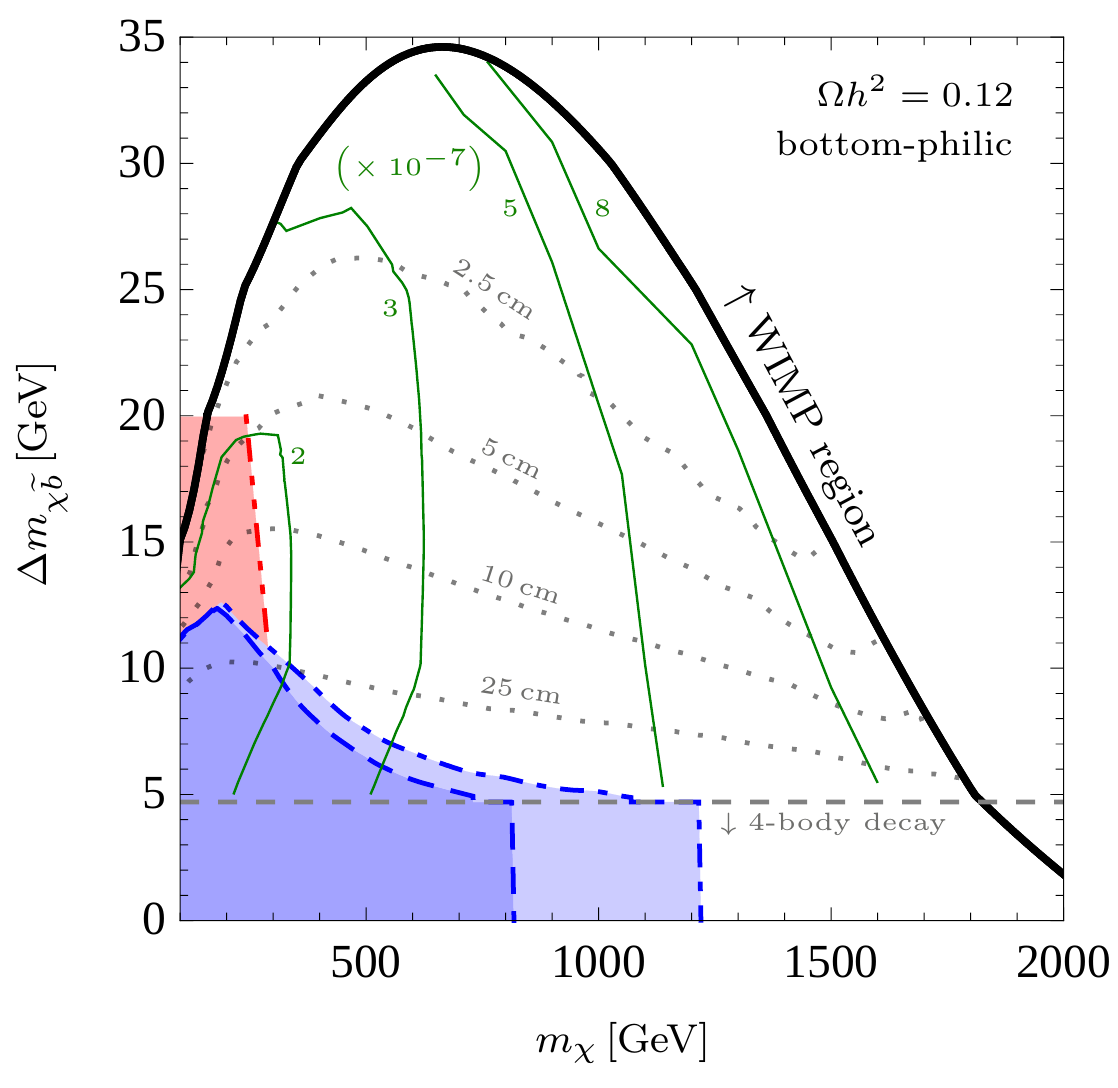}}
\put(0.505,-0.02){\includegraphics[width=0.4788\textwidth]{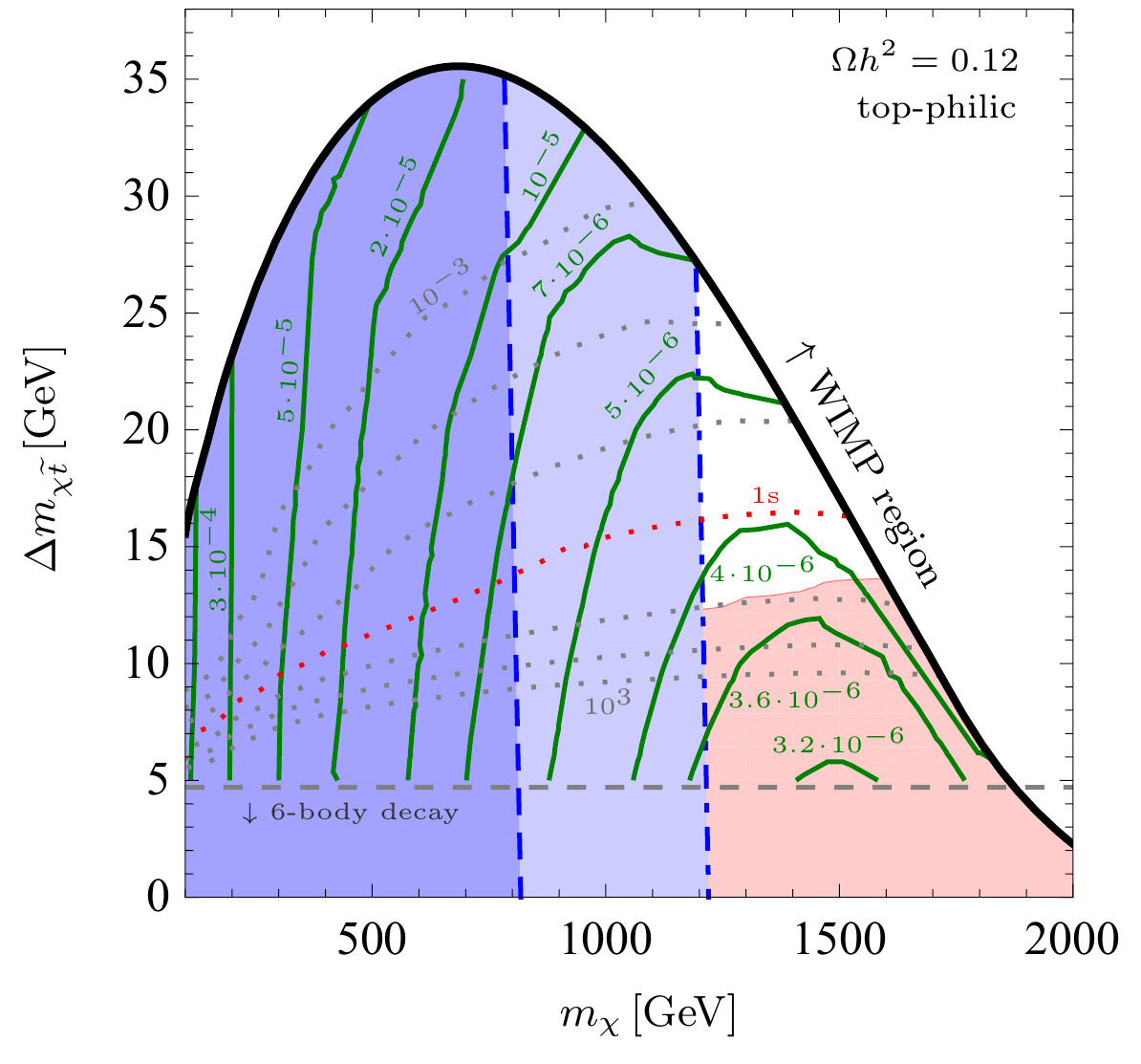}}
        \end{picture}
    \caption{
Cosmologically viable parameter space ($\Omega h^2=0.12$)
in the conversion-driven freeze-out region (below black thick curve) for the bottom- (left) and top-partner (right) mediator.$^{\!3,4}$
Contours of constant $\lambda_\chi$ are shown in green ($\times 10^{-7}$ in the left plot).
Contours of constant mediator decay length (left) and lifetimes (right) are shown as gray dotted curves.
The displayed lifetimes range from $10^{-3}\,$s to $10^{3}\,$s in steps of an order of magnitude (the curve for 1\,s is highlighted in red for better readability). The 95\% C.L. exclusion regions from $R$-hadron searches at the 8 and 13\,TeV LHC are shown in dark and light blue, respectively. The red shaded region bordered by the red dot-dot-dashed curve in the left plot denotes the constraint from monojet searches at the 13\,TeV LHC whereas the light red shaded region in the right plot denotes constraints from BBN\@. Below the horizontal gray dashed line ($\sim 5\,$GeV) the 2- (left) and 4-body decay (right) is kinematically forbidden rendering the 4- and 6-body decay, respectively, to be dominant.
    }
    \label{fig:nonCEcont}
\end{figure}

Without the assumption of chemical equilibrium between dark matter and the mediator the computation of the relic density
requires the solution of the full coupled set of Boltzmann equations explicitly including conversion processes in the dark sector. We take into account the leading decay process (2- and 4-body decay for the bottom- and top-philic model, respectively) and all $2\to2$ scattering processes (as well as the leading $2\to3$ processes for the top-philic model).\cite{Garny:2017rxs,Garny:2018icg}
The cosmologically viable parameter space ($\Omega h^2=0.12$)\,\cite{Ade:2015xua} is shown in Fig.~\ref{fig:nonCEcont} for the bottom- (left) 
and top-philic (right) model in the plane spanned by the dark matter mass and the mass difference $\Delta_{\chi \tilde q}=m_{\tilde q}-m_\chi$. 

\begin{figure}[h]
\centering
\setlength{\unitlength}{1\textwidth}
\begin{picture}(0.45,0.435)
\put(-0.01,-0.02){\includegraphics[width=0.45\textwidth]{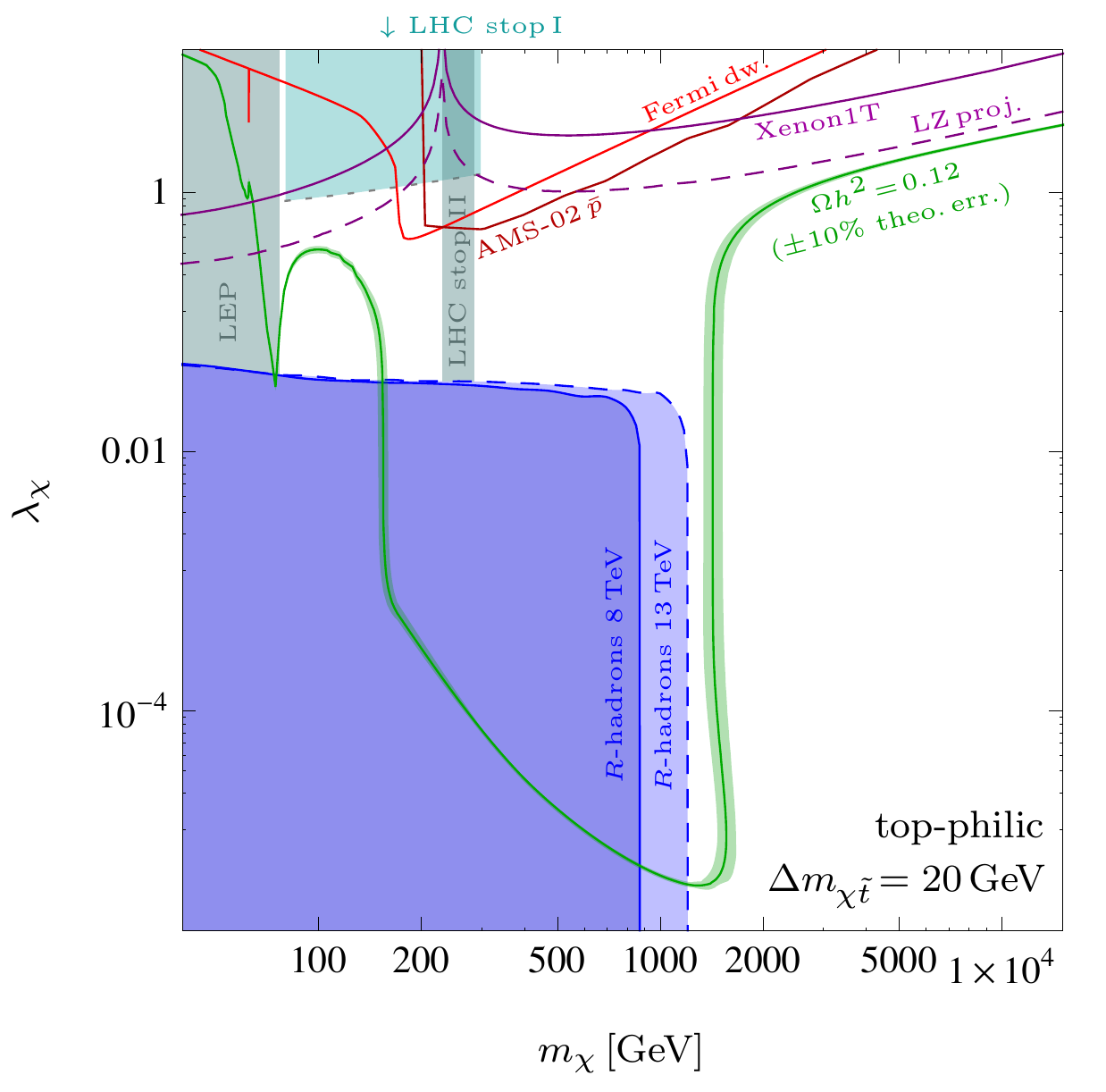}}
\end{picture}
\caption{
Constraints on the coupling $\lambda_\chi$ 
as a function of the dark matter mass for $\Delta m_{\chi \tilde t}=20\,\mathrm{GeV}$
in the top-philic model.$^{\!4}$
The green curve and green shaded band shows the
coupling that provides $\Omega h^2=0.12$ and its theoretical 
uncertainty, respectively, assuming a relative error of 
10\% on the prediction for $\Omega h^2$.
The 95\% C.L. exclusion regions from $R$-hadron searches at the 8 and 13\,TeV LHC are 
shown in dark and light blue, respectively. For comparison, we show limits from canonical 
WIMP searches, \emph{i.e.}~95\% C.L. upper limits on $\lambda_\chi$ from indirect detection searches from
Fermi-LAT dwarfs (light red curves) and AMS-02 antiprotons (dark red curves)
as well as 90\% C.L. direct detection upper limits from Xenon1T 2017 (purple solid curve)
and direct detection projections for the LZ experiment (purple dashed curves). 
Additionally, we show 95\% C.L. exclusion regions from searches for supersymmetric top-partners at 
the LHC and LEP\@. Further details can be found in Ref.~4.
}
\label{fig:1Dslice}
\end{figure}

Above the black thick curve relative chemical equilibrium holds resembling a standard WIMP/coannihilation scenario
while below this curve solutions for conversion-driven freeze-out exist where $\lambda_\chi$ is in the range $10^{-6}$--$10^{-7}$ and
$10^{-3}$--$10^{-6}$ for the bottom- and top-philic model, respectively. At the curve itself the measured relic density can be
obtained for a wide range of $\lambda_\chi$ that provide a negligible dark matter annihilation cross section but still sizable conversion rates
maintaining chemical equilibrium in the dark sector. For illustration, Fig.~\ref{fig:1Dslice} shows the respective solution for $\lambda_\chi$ as a function of the dark matter mass for the top-philic model and for a fixed mass splitting of $\Delta m_{\chi \tilde t}=20\,\mathrm{GeV}$. At the transition from the WIMP to the conversion-driven freeze-out region the coupling drops by several orders of magnitude.

As discussed in Sec.~\ref{sec:LLPmir} conversion-driven freeze-out 
predicts macroscopic decay length of the heavier $Z_2$-odd state. For the bottom-partner the 2-body decay of the mediator is open rendering both conversion via decays and scatterings to be similarly important and hence $\Gamma_\mathrm{dec}\sim H$ during freeze-out. As a consequence the decay length is of the order of several cm, see gray dotted curves in the left panel of Fig.~\ref{fig:nonCEcont}. For the top-philic model, in the parameter region of interest, the leading decay channel is a 4-body decay. Accordingly, conversions during freeze-out are mediated solely by scatterings while the decay becomes efficient only well after freeze-out. Note that mediator lifetimes above ${\cal O} (1\,\mathrm{s})$ are subject to constraints from big bang nucleosynthesis (BBN)\,\cite{Jedamzik:2006xz} indicated by the 
red shaded region in the right panel of Fig.~\ref{fig:nonCEcont}.

\subsection{Constraints from LLP searches}

At the LHC mediators pairs could be copiously produced. Being a colored state the mediator is expected to hadronize and form $R$-hadron bound states. Depending on its decay length, it will typically decay inside
or traverse the detector. 
Accordingly, for the case of the bottom-partner the signatures of kinked or disappearing tracks (depending on the detectability of the radiated $b$-quark) provide promising search channels. Although similar searches have been
performed for supersymmetric models\,\cite{Aad:2015rba,Aad:2013yna,CMS:2014gxa} these cannot be reinterpreted 
within the present model without additional information provided by the collaborations. For instance, searches for 
disappearing tracks are performed under the assumption of purely electrically charged particle while $R$-hadron 
undergo a more complicated traverse through the detector being able to flip charge or become neutral through interactions with the detector material. Therefore their applicability is unclear.

Searches for detector-stable $R$-hadrons can, however, be reinterpreted for finite decay lengths\,\cite{Garny:2017rxs} 
using the signature efficiencies provided for heavy stable charge particle searches released by CMS.\cite{Khachatryan:2015lla}
The resulting constraints obtained from the 8\,TeV~\cite{Chatrchyan:2013oca}
and 13\,TeV~\cite{CMS-PAS-EXO-16-036} LHC data are shown in Fig.~\ref{fig:nonCEcont}
as the dark and light blue shaded regions, respectively. 
Note that the model can also be constrained by mono-jet searches exploiting a large missing energy from initial state radiation in the mediator production process. Here we show the limit from ATLAS using $3.2\,\text{fb}^{-1}$ of 13\,TeV data,\cite{Aaboud:2016tnv} see the red shaded region in the left panel of Fig.~\ref{fig:nonCEcont}.

While a dedicated search is expected to be able to significantly increase the sensitivity to the bottom-philic model, 
the top-philic model -- featuring a detector-stable mediator 
 -- is already very well constraints from $R$-hadron searches. Dedicated searches for meta-stable top-partners could, however, fill some gaps in the sensitivity outside the conversion-driven freeze-out region where the mediator tends to have intermediate lifetimes, \emph{cf.}~Fig.~\ref{fig:1Dslice}.
Note that the entire cosmologically allowed conversion-driven freeze-out region
of the top-philic model is expected to be probed by $R$-hadron searches at 13\,TeV
with an integrated luminosity of approximately $300\,\text{fb}^{-1}$.\cite{Garny:2018icg}

\section{SuperWIMPs and freeze-in}\label{sec:sW}

So far we have considered scenarios where dark matter undergoes a phase of thermalization and freeze-out.
Another possibility is that dark matter never reaches chemical equilibrium with the standard model bath
being produced through out-of-equilibrium processes. 
While we cannot hope to observe an entirely 
thermally decoupled dark sector,\cite{Kahlhoefer:2018xxo} a partly thermalized dark 
sector can provide promising prospects to be explored at the LHC\@. There are two main scenarios for non-thermalized dark matter genesis in the literature: the freeze-in\,\cite{Hall:2009bx} 
and the superWIMP\,\cite{Feng:2003uy} scenario. While freeze-in mediated by
a $Z_2$-even 
mediator in general does not provide promising prospects for the LHC,\footnote{A valuable exception can, however, 
arise from a non-standard cosmological history.\cite{Co:2015pka}} both scenarios may be observable, if dark matter production is mediated by a $Z_2$-odd state.\footnote{Here we concentrate on dark matter with masses in the GeV to TeV range, typically providing detector-stable mediators in the regime of thermally decoupled dark matter while lighter dark matter can also provide decays within the LHC detector, \emph{e.g.}~displayed vertices.\cite{Calibbi:2018fqf}}

\begin{figure}[t]
\centering
\setlength{\unitlength}{1\textwidth}
\begin{picture}(0.45,0.44)
\put(-0.03,-0.0117){\includegraphics[width=0.48\textwidth]{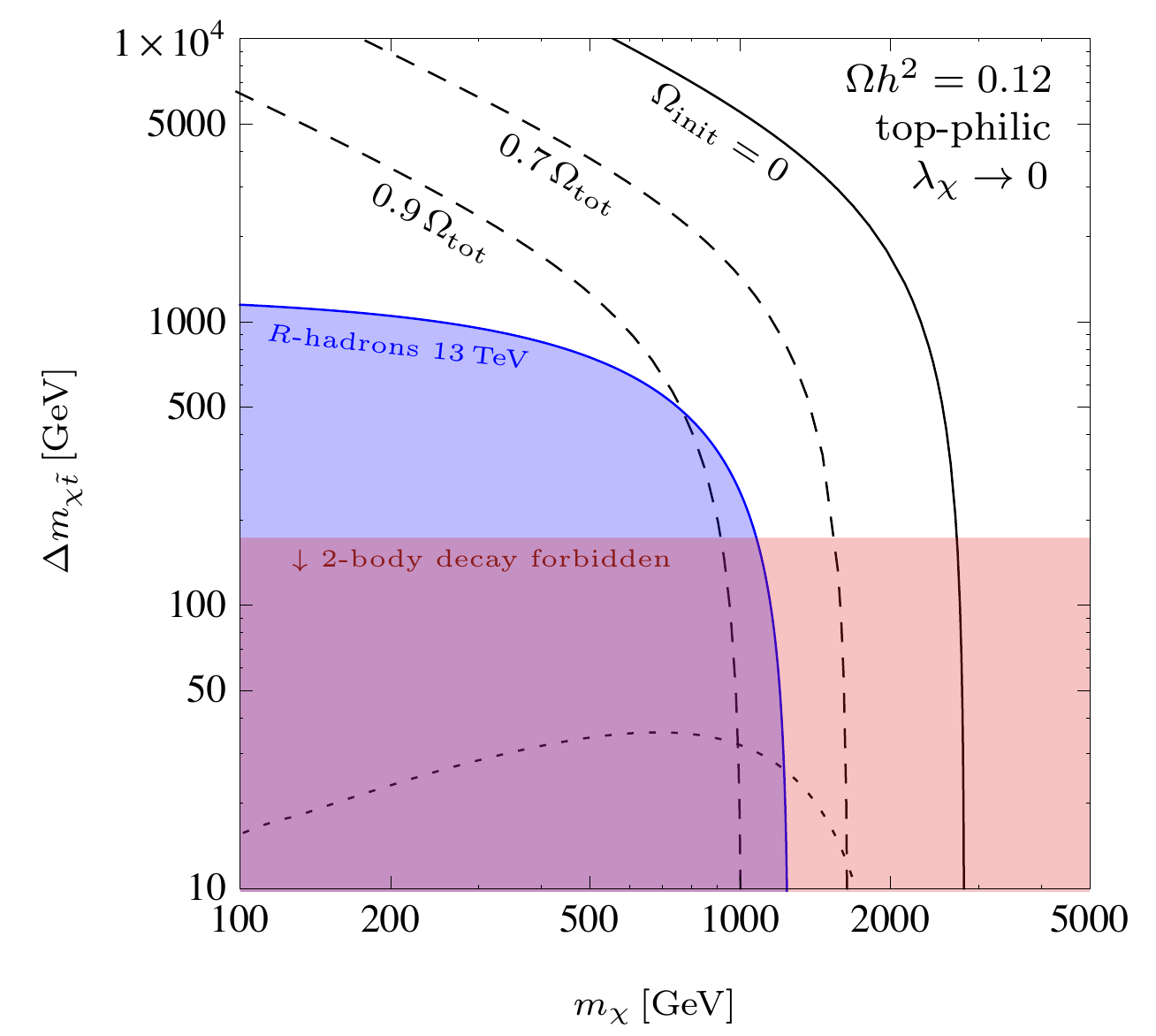}}
\end{picture}
\caption{
Cosmologically viable parameter space ($\Omega h^2=0.12$)
for a superWIMP scenario ($\lambda_\chi\to0$). The solid as well as the dashed black lines show 
the corresponding slices in parameter space for three different choices of the abundance prior to 
the decay of the mediator: a vanishing abundance as well as a fraction of 0.7 and 0.9, respectively, of the total
(\emph{i.e.}~final) abundance. The light blue shaded area denotes the 95\% C.L. exclusion region from $R$-hadron 
searches at the 13\,TeV LHC\@. For comparison, the dotted black curve shows the border of the conversion-driven
freeze-out region. 
}
\label{fig:sW}
\end{figure}

In the simplified model introduced in Eq.~\eqref{eq:tchmodel} the leading 
contributions to dark matter production would be the conversion processes $\tilde q i \leftrightarrow \chi  j$ and $\tilde q \leftrightarrow \chi  j$, where $i,j$ denote standard model particles. Depending
on the masses, coupling strength and the model under consideration ($q=b,t$) the dominant contribution to 
dark matter production occurs around or after the freeze-out of $\tilde q$. The first case constitutes a realization 
of freeze-in while the second case resembles a superWIMP scenario. However, in general both contributions are present. Furthermore, as the model is sensitive to the initial conditions -- which are not washed out by thermalization -- a further contribution might stem from physics relevant at earlier times that are not captured by the (low-energy) simplified model considered here, \emph{e.g.}~a contribution from reheating.

In Fig.~\ref{fig:sW} we plot the cosmologically viable viable parameter space ($\Omega h^2=0.12$)
for the superWIMP scenario for the top-philic model ($q=t$). We assume a sufficiently small coupling so that 
the mediator decay occurs well after its freeze-out and plot the curve for which 
\begin{equation}
(\Omega h^2)_\chi = \frac{m_\chi}{m_{\stop}}\, (\Omega h^2)_{\stop} + (\Omega h^2)_\chi^\mathrm{init}=0.12\,,
\end{equation}
where $(\Omega h^2)_{\stop}$ is the freeze-out abundance of the mediator in the absence of any coupling to dark matter while
$(\Omega h^2)_\chi^\mathrm{init}$ is the initial dark matter abundance prior to the mediator decay. We show the result for three different choices of $(\Omega h^2)_\chi^\mathrm{init}$: a vanishing abundance as well as a fraction of 0.7 and 0.9, respectively, of the total
(\emph{i.e.}~final) abundance. This initial abundance represents a possible contribution from the very early Universe (\emph{e.g.}~reheating phase) or from freeze-in. In the model under consideration the latter would arise from conversion via scattering and could be computed for a given $\lambda_\chi$. 

The red shaded region in Fig.~\ref{fig:sW} denotes a mass splitting below the top mass, introducing an additional suppression
of the mediator decay leading to large lifetimes that are potentially 
in conflict with BBN for the very small couplings considered here.
However, outside this region where the 2-body decay of the mediator is open we find that consistency with BBN
can easily be achieved in the limit of a dominant superWIMP scenario, 
\emph{e.g.}~$\tau_{\st} <1\,$s requires $\lambda_\chi \gtrsim10^{-12}$ 
for which conversion rates are entirely inefficient, $\Gamma_\text{conv}\ll H $, until well after the freeze-out
of the mediator.
The blue shaded region shows the respective limit for detector-stable mediators from $R$-hadron searches at the $13\,$TeV LHC\@.\cite{CMS-PAS-EXO-16-036}
It constraints part of the parameter space with relatively large $(\Omega h^2)_\chi^\mathrm{init}$ where the mediator freeze-out abundance, $(\Omega h^2)_{\stop}$, is required to be small. For comparison we show the boundary of the conversion-driven freeze-out as a dotted black curve.

\section{Conclusions}\label{sec:Con}

In this article we discussed dark matter scenarios that predict long-lived particles at the LHC\@. We focussed on models with a $Z_2$-odd dark sector providing three distinct regions characterized by a decreasing coupling strength: The well-know coannihilation scenario, conversion-driven freeze-out and the superWIMP/freeze-in scenario. The latter two cases exploit an intrinsic connection between the dynamics of dark matter genesis and the involved decay rates of a heavier $Z_2$-odd state. For particles in the GeV to TeV range a departure from relative chemical equilibrium during its freeze-out implies macroscopic decay length at the LHC -- an intriguing coincidence that renders the LHC to be a powerful tool to explore these scenarios.

We presented realizations of conversion-driven freeze-out
within the framework of simplified dark matter with a top- and bottom-partner mediator.
While the former model predicts detector-stable $R$-hadrons that are well constrained by existing searches at the LHC, the latter provides decay length of the order of several cm.  These are relatively poorly constraint as existing searches for meta-stable colored states cannot be straightforwardly reinterpreted within the model. Dedicated searches are required for a full exploration of the model.
Despite the small couplings required by conversion-driven freeze-out ranging between $10^{-3}$ and $10^{-7}$ it provides an efficient thermalization of dark matter prior to freeze-out. This is contrast to the superWIMP scenario where dark matter never reaches thermal equilibrium with the standard model bath. 
The cosmologically viable parameter space, hence, depends on a possible contribution of dark matter production from the very early Universe as well as from freeze-in around the time of the mediator freeze-out and contains a wide range of yet unexplored masses. Parts of the parameter space can be probed with $R$-hadron searches at the 13 TeV LHC\@.

\section*{Acknowledgements}

This work is supported by the German Research Foundation DFG through the 
research unit ``New physics at the LHC''.

\end{document}